\documentclass[year=26]{fmcad}
\usepackage{graphicx}
\usepackage[section]{placeins}
\usepackage{amsmath}

\title{Formal Performance and Compile Time Guarantees for Compiler Optimization Heuristics}

\author{\IEEEauthorblockN{Nikil V. Shyamsunder}
\IEEEauthorblockA{
Department of Computer Science, Cornell University \\
nvs26@cornell.edu}}

\begin{document}

\maketitle
\begin{abstract}
Modern optimizing compilers rely on heuristic search algorithms for NP-hard optimization problems, which can result in poor generated-code performance and long or unpredictable compile times. These are considered bugs by users, but verified compilers rarely reason beyond semantic preservation. We propose verifying performance and compile time properties of compiler passes. As a proof-of-concept, we formulate inline expansion using a cost model estimating instruction-cache performance. We mechanize this in Rocq, prove semantic preservation of the inlining transformation, and verify the algorithm's monotone improvement, convergence-time bound, and performance bounds for intermediate and final solutions.
\end{abstract}

\section{Introduction}

Compilers are evaluated on correctness, the performance of generated code, and
the compile time required to achieve it. Verified compilers have traditionally focused exclusively on correctness. CompCert, for example, leverages the technique of \emph{bisimulation} proofs for semantic preservation: a source program step must correspond to a sequence of target steps with the same behavior, and vice versa \cite{leryoycomp}. Recent extensions use \emph{quantitative} reasoning to guarantee resource availability on the target machine, such as stack-space or memory bounds \cite{veristack,myreen2021cakeml}. These rule out a large class of miscompilations, but do not reason about the optimality of the algorithms: how close a pass's output is to the best result in its search space.

In practice, excessively poor performance of generated code and long or unpredictable compile times are also treated as compiler bugs, revealing a mismatch between what verification guarantees and what users expect from a compiler. These bugs are often caused by global
optimization passes where compilers rely on heuristics, such as register allocation and inline expansion in traditional software compilers \cite{theodoridis2022optimalinlining, chaitin1981register, poletto1999linear}.
A heuristic may find an excessively suboptimal solution or search too long for an acceptable solution.
This problem is magnified in compilers for less conventional targets, such as ML frameworks and hardware toolchains, where the abundance of NP-hard problems such as tensor layout, hardware scheduling, placement, and routing means a poor search heuristic can miss high-performance transformations or incur long compile times \cite{zheng2020ansor, sahni1980complexity, drabek2002list, cong2006efficient,
jain2017resource}.
In all of these settings, the compiler runs without formal guarantees that search will conclude
quickly or find an effective solution.
%
% Moreover, poor decisions in one pass, such as inlining, layout, or allocation, can amplify code-size growth or restrict the
% optimization opportunities available to later passes.

For an entire production compiler, generated-code quality and compile time are
emergent properties of many interacting passes and cost models, making them difficult to state as a single specification.
Instead, we aim for guarantees about the particular passes where poor generated-code
quality and runaway compile time are most likely to arise. Specifically, we focus on verifying two kinds of
properties:
\begin{itemize}
\item \textbf{Performance bounds.} To rule out excessively poor generated code, we want a machine-checked bound on a pass's output relative to the best solution in its search space under some cost model. Even when no such bound is found, the formal reasoning is itself informative: it tends to pinpoint the input configurations under which the pass degrades, exposing the areas worth ameliorating.
\item \textbf{Convergence time and progress bounds.} We want guarantees
that a compiler optimization terminates in a bounded number of steps, and that
varying the level of optimization through controls like \texttt{-O\{N\}} flags gives predictable compile time/code-performance
tradeoffs.
\end{itemize}
We report our initial findings on a simple pass and cost model, as a first step toward providing production-ready algorithms that are provably correct and performant for critical passes.
%
% \vspace{-7.5pt}
\section{Modeling Inline Expansion}
\label{model}

As an initial example, we use \emph{inline expansion}, which replaces a
function call with a copy of the callee body. Inlining can remove call overhead and expose further optimizations, but may increase compile time, code size, and instruction-cache (i-cache) pressure. Inline expansion is at least as hard as the knapsack problem under a simple model where each candidate call site has an independent code size cost and expected benefit \cite{scheifler1977inline}. Standard knapsack approximation algorithms can therefore be used, but the
independence assumption is strong since inlining one function changes the
cost and profitability of later call sites. Heuristic approaches
have been proposed using code size, i-cache
behavior, optimization potential, call-graph ordering, and profile information
\cite{scheifler1977inline,mcfarling1991procedure,chakrabarti2006inline,serrano1997inline,zhao2003inline,theodoridis2022optimalinlining}.
LLVM's inliner uses many of the above approaches, iterating over a worklist of call-graph components and processing call sites in bottom-up order. After a successful inlining, costs are recomputed and newly exposed call sites are appended.

For our prototype, we base our pass on LLVM's inliner but make certain simplifications. First, we model only i-cache performance costs. Formally, let $\mathcal{P}$ be the set of decisions the optimization must make;
for inline expansion, each element $i \in \mathcal{P}$ is a candidate call site. Each call site
chooses from $\mathcal{S}_i = \{\texttt{inline}, \texttt{preserve}\}$, and
$\mathcal{S}$ is the space of choices for all call sites. Let $\mathcal{R}$ be the set of
shared i-cache-line resources. For each call site $i$, there is an arbitrary mapping $\rho_i : \mathcal{S}_i \to 2^{\mathcal{R}}$ from each choice to the
resources it uses. The cost function $\mathcal{C}$
encodes resource contention: performance is worse when more selected call sites
share the same cache line, which we denote as \emph{execution delay}. Thus, we model each
resource $r$ with an affine delay cost $c_r(x) = a_r x + b_r$, where $x$ is the
number of selected call sites using that resource. For an inlining solution profile $s \in
\mathcal{S}$, a call site's delay $D_i$ and the total execution delay for the program $\mathcal{C}$ are
\[
D_i(s) = \sum_{r \in \rho_i(s_i)} c_r(|\{j : r \in \rho_j(s_j)\}|),
\qquad
\mathcal{C}(s)=\sum_{i \in \mathcal{P}} D_i(s).
\]
% \vspace{-2pt}
This abstracts away many hardware details, but captures
the first-order fact that inlined code increases code size and can raise pressure
on hot i-cache lines.

Second, we assume all candidates are
\emph{leaf} call sites, so inlining a candidate does not introduce additional
candidate calls into the worklist. Rather than a single iteration, we generalize by allowing multiple passes over the call sites in \emph{any} ordering and run until reaching a fixed point. Specifically, each call site
evaluates its execution delay under its two available choices and flips from
\textit{preserve} to \textit{inline}, or vice versa, exactly when the other
choice strictly reduces its delay; the procedure repeats these locally
improving steps, recomputing costs at each step, until no candidate can improve. We now bound the worst-case convergence time to a fixed point, and the worst-case performance of the intermediate and final solutions.
% \vspace{-4pt}
\section{Convergence \& Performance Guarantees}
\label{informal}
We begin by noting that our inlining procedure is an instance of a
well-studied mathematical structure from game theory. Treating call sites as players, their two
choices as strategies, and their delays as payoffs (or in our case costs) gives a strategic
game~\cite{roughgardenAGT}; more specifically, our choices of $\mathcal{P}$,
$\mathcal{S}$, and $\mathcal{C}$ form a \emph{congestion
game}~\cite{rosenthal1973} over the resource set $\mathcal{R}$. Our fixed-point iteration procedure is precisely \emph{iterative best response} in game theory: repeatedly let one player switch to its best selfish choice while holding the others fixed~\cite{roughgardenAGT}.

% This identification lets us reuse proof strategies and results from algorithmic
% game theory, many of which already have proof-assistant formalizations in Rocq,
% Lean, and Isabelle/HOL \cite{bagnall2017library, garg2026econcslib, ramos2026nash}.

Congestion games belong to the class of \emph{exact
potential games}~\cite{monderer1996potential}, where a single global
\emph{potential function} reflects each individual player's cost change exactly~\cite{roughgardenAGT}. The exact-potential property guarantees that every selfish, \emph{local} inlining
decision made via iterative best response strictly decreases a \emph{global} progress
measure. Consequently, the
pass is guaranteed to converge to a fixed point, which is a \emph{pure Nash equilibrium} (PNE) where no player can unilaterally improve. At a PNE, we can directly verify our performance bound by leveraging the \emph{Price of Anarchy} (PoA). The PoA is the ratio of the maximum value of $\mathcal{C}$ at any PNE to the global minimum value of $\mathcal{C}$.
% $$PoA = \frac{\max_{s \in PNE} \mathcal{C}(s)}{\min_{s \in \mathcal{S}} \mathcal{C}(s)}$$

Because our specific formulation models i-cache contention using linear delays, it acts as an \emph{affine} congestion game, for which Christodoulou and Koutsoupias bound the PoA at 2.5: our algorithm is at worst $2.5\times$ the global optimum \cite{christodoulou2005price}.
% \[
% 2 \cdot \mathsf{Cost}({\mathsf{Nash}})
%   \le 5 \cdot \mathsf{Cost}({\mathsf{OPT}}).
% \]
While quite loose, this bound holds for any linear cost model over any cache geometry and code layout. This leaves a clear path for future work to tighten the bound by incorporating the specific details of the target architecture and program structure.

We also obtain an explicit compile time progress bound. We work with
natural-valued costs, so each accepted best-response step decreases the
underlying global potential function by at least one. If there are $N$ candidate
call sites, $E$ cache-line resources, and every resource cost satisfies
$a_r \le A$ and $b_r \le B$, then this progress measure is initially bounded by
$E\,(A N^2 + B N)$. Therefore iterative best response converges within
$E\,(A N^2 + B N)$ accepted steps. More generally, after any $k$ accepted steps,
the same argument shows that the remaining number of steps before convergence is at
most $E\,(A N^2 + B N)-k$. We also prove 
 $ \mathcal{C}(s_k) \le 2(E(AN^2+BN)-k)$, yielding a coarse intermediate performance bound. This gives a predictable
compile time/code-performance tradeoff, which \texttt{-O\{N\}} levels can expose: in a ``fast compilation'' mode like \texttt{-O1}, the compiler might halt the pass after a fixed $k$ steps, and users can reason about the resulting performance.

\vspace{-1pt}
\section{Mechanization in Rocq}

In order to validate our methods, we mechanize our prototype in Rocq over a
language inspired by CompCert's RTL, an intermediate representation where
functions are control-flow graphs over elementary instructions that read and
write pseudo-registers. We define the language's syntax and small-step
operational semantics and define
leaf inline expansion as a source-to-source relation parameterized by an
arbitrary candidate call site. We then prove, under bisimulation, that the
transformation defined by this relation is semantics-preserving.

Next, we formalize the iterative-best-response procedure that uses the affine cost model to decide \emph{which} candidate call sites to inline. Once a
subset of call sites has been chosen by the procedure, the
compiler ``performs'' those inlinings using the expansion relation. We mechanically
verify the PoA and convergence guarantees detailed in \ref{informal}. The definitions and proofs are
approximately $4{,}000$ LOC with 0 admitted lemmas.

\section{Conclusion \& Future Work}
We show that a verified compiler pass \emph{can and should} certify performance bounds and convergence-time properties. Although this prototype is especially amenable to a strategic-game formulation and has strong theoretical guarantees due to its potential-game structure, we expect the same proof strategy to extend to richer cost models and other optimization passes, since game theory provides a natural language for relating local heuristic decisions to the global optimum and may allow us to reuse existing mechanized results \cite{bagnall2017library, garg2026econcslib, ramos2026nash}. Traditional approximation algorithm proof techniques may be required in certain settings. Three directions remain. First, we aim to instantiate the cost model with concrete, architecture-specific details, and to enrich the cost model toward production inliners such as LLVM's. Second, we plan to model \emph{joint} cost models across multiple passes, which capture how one pass affects downstream behavior,
enabling compositional guarantees that address phase ordering directly.
Finally, we aim to extend the approach to more compelling optimization passes with unpredictable performance and compile time properties such as tensor layout or instruction scheduling.

\pagebreak
\bibliographystyle{IEEEtran}
\bibliography{refs}
\end{document}